\documentclass[aps,prb,twocolumn,groupedaddress]{revtex4}

\usepackage{graphics}
\usepackage{graphicx}

\begin{document}

\title{Tuning the magnetism of ordered and
disordered
strongly-correlated \\
 electron nanoclusters}
\author{Nicholas Kioussis, Yan Luo, and Claudio Verdozzi}
\email[E-mail me at: ]{yan.luo@csun.edu. } \affiliation{Department
of Physics, California State University Northridge, California
91330-8268}

\begin{abstract}
Recently, there has been a resurgence of intense experimental and
theoretical interest on the Kondo physics of nanoscopic and
mesoscopic systems due to the possibility of making experiments in
extremely small samples. We have carried out exact diagonalization
calculations to study the effect of energy spacing $\Delta$ in the
conduction band states, hybridization, number of electrons, and
disorder on the ground-state and thermal properties of
strongly-correlated electron nanoclusters. For the ordered
systems, the calculations reveal for the first time that $\Delta$
tunes the interplay between the {\it local} Kondo and {\it non
local} RKKY interactions, giving rise to a ``Doniach phase
diagram" for the nanocluster with regions of prevailing Kondo or
RKKY correlations. The interplay of $\Delta$ and disorder gives
rise to a $\Delta$ versus concentration $T=0$ phase diagram very
rich in structure. The parity of the total number of electrons
alters the competition between the Kondo and RKKY correlations.
The local Kondo temperatures, $T_K$, and RKKY interactions depend
strongly on the local environment and are overall {\it enhanced}
by disorder, in contrast to the hypothesis of ``Kondo disorder''
single-impurity models. This interplay may be relevant to
experimental realizations of small rings or quantum dots with
tunable magnetic properties.
\end{abstract}

\maketitle

\section{INTRODUCTION}
Magnetic impurities in non-magnetic hosts have been one of the
central subjects in the physics of strongly correlated systems for
the past four decades\cite{Hewson}. Such enduring, ongoing
research effort is motivated by a constant shift and increase of
scientific interest over the years, from dilute \cite{1960} to
concentrated impurities \cite{Steglich}, from periodic
\cite{UedaRMP} to disordered samples \cite{Miranda, CastroNeto},
and from macroscopic \cite{Stewart} to nanoscale phenomena
\cite{Schlottman}. Macroscopic strongly correlated electron
systems at low temperatures and as a function of magnetic field,
pressure, or alloying show a wide range of interesting phenomena,
such as non-Fermi-liquid behavior, antiferromagnetism,
ferromagnetism, enhanced paramagnetism, Kondo insulators, quantum
criticality or superconductivity\cite{Hewson, Stewart}. These
phenomena are believed to arise through the interplay of the Kondo
effect, the electronic structure and intersite correlations.  In
the simplest single-impurity case, the Kondo problem describes the
antiferromagnetic interaction, $J$, between the impurity spin and
the free electron spins giving rise to an anomalous scattering at
the Fermi energy, leading to a large impurity contribution to the
resistivity\cite{Hewson}. The low-energy transport and the
thermodynamic properties scale with a single characteristic
energy, the Kondo temperature, $T_K \propto exp(-1/\rho(E_F)J)$,
where $\rho(E_F)$ is the density of states of the conduction
electrons at the Fermi energy \cite{Hewson}. At $T
>> T_K$, the impurity spin is essentially free and  the problem
can be treated perturbatively. At $T<<T_K$, the impurity spin is
screened forming a singlet complex with the conduction electrons,
giving rise to a local Fermi liquid state.

For dense Kondo or heavy fermion compounds containing a periodic
array of magnetic ions interacting with  the sea of conduction
electrons, the physics is determined from the competition between
the {\it non local} Ruderman-Kittel-Kasuya-Yosida (RKKY)
interactions and the {\it local} Kondo interactions\cite{Si}. The
RKKY interaction is an indirect magnetic interaction between
localized moments mediated by the polarized conduction electrons,
with an energy scale of order $J_{RKKY} \propto J^2\rho(E_F)$,
which promotes long- or short-range magnetic ordering. On the
other hand, the Kondo effect favors the formation of singlets
resulting in a non-magnetic ground state. In the high temperature
regime the localized moments and the conduction electrons retain
their identities and interact weakly with each other. At
low-temperatures, the moments order magnetically if the RKKY
interaction is much larger than the Kondo energy, while in the
reverse case, the system forms a heavy Fermi liquid of
quasiparticles which are composites of local moment spins bound to
the conduction electrons\cite{Stewart, Si}. Thus, the overall
physics can be described by the well-known ``Doniach phase
diagram", originally derived for the simple Kondo necklace
model\cite{doniach}. The description of the low-temperature state,
when both the RKKY and the Kondo interactions are of comparable
magnitude, is an intriguing question that remains poorly
understood and is the subject of active research\cite{Si}.

The interplay of disorder and strong correlations has been a
subject of intensive and sustained research, in view of the
non-Fermi-liquid (NFL) behavior in the vicinity of a quantum
critical point\cite{schroder}.
 Various disorder-driven models have been proposed to explain the
experimentally observed\cite{Stewart} NFL behavior at low
temperatures\cite{Miranda, CastroNeto, Stewart, Riseborough}. The
phenomenological ``Kondo disorder'' approaches \cite{Miranda,
Bernal}, based on single-impurity models, assume  a distribution
of Kondo temperatures caused by a distribution of either $f-c$
orbital hybridization or of impurity energy levels. These models
rely  on the presence of certain sites with very low $T_K$ spins
leading to a NFL behavior at low $T$. An open issue in such
single-site Kondo approaches,  is whether the inclusion of RKKY
interactions would renormalize and eliminate the low-$T_K$
spins\cite{UedaRMP, rice, read, assaad}. An alternative view is
the formation of large but finite magnetic clusters (Griffith
phases) within the disordered phase through the competition
between the RKKY and Kondo interactions \cite{CastroNeto,
Miranda2001}.

On the other hand, the possibility of making experiments in
extremely small samples has lead to a resurgence of both
experimental and theoretical interest of the physics of the
interaction of magnetic impurities in nanoscopic and mesoscopic
non-magnetic metallic systems. A few examples include quantum
dots\cite{gordon}, quantum boxes\cite{thimm} and quantum
corrals\cite{manoharan}. Recent scanning tunneling microscope(STM)
experiments\cite{odom} studied the interaction of magnetic
impurities with the electrons of a single-walled nanotube confined
in one dimension. Interestingly, in addition to the bulk Kondo
resonance new sub peaks were found in shortened carbon nanotubes,
separated by about the average energy spacing, $\Delta$, in the
nanotube.  The relevance of small strongly correlated systems to
quantum computation requires understanding how the infinite-size
properties become modified at the nanoscale, due to the  finite
energy spacing $\Delta$ in the conduction band which depends on
the size of the particle \cite{ Schlottman, thimm, Hu, Balseiro,
Affleck}. For such small systems, controlling $T_K$ upon varying
$\Delta$ is acquiring increasing importance since it allows to
tune the cluster magnetic behavior and to encode quantum
information. While the effect of $\Delta$ on the single-impurity
Anderson or Kondo model has received considerable
theoretical\cite{Schlottman, thimm, Hu, Balseiro, Affleck} and
experimental\cite{odom} attention recently, its role on {\it
dense} impurity clusters remains an unexplored area thus far. The
low-temperature behavior of a nanosized heavy-electron system was
recently studied within the mean-field
approximation\cite{schlottmann2001}. A central question is what is
the effect of $\Delta$ on the interplay between the Kondo effect
and the RKKY interaction

In this work we present exact diagonalization calculations for
$d$- or $f$-electron nanoclusters to study the effects of energy
spacing, parity of number of electrons, and hybridization on the
interplay between Kondo and RKKY interactions in both {\it
ordered} and {\it disordered} strongly correlated electron
nanoclusters. While the properties of the system depend on their
geometry and size\cite{Pastor}, the present calculations treat
exactly the Kondo and RKKY interactions, the disorder averages,
and they provide a distribution of local $T_K$'s renormalized by
the intersite f-f interactions. Our results show that: i) tuning
$\Delta$ and the parity of the total number of electrons can drive
the nanocluster from the Kondo to the RKKY regime, i.e. a zero-
temperature energy spacing versus hybridization phase diagram; ii)
the temperature versus hybridization ``Doniach" phase diagram for
nanoclusters depends on the energy spacing ; iii)  changing the
total number of electrons from even to odd results in an
enhancement (suppression) of the local Kondo (RKKY) spin
correlation functions; iv) the $\Delta$ versus alloy concentration
$T=0$ phase diagram exhibits regions with prevailing Kondo or RKKY
correlations alternating with domains of ferromagnetic (FM) order;
and v) the local $T_K$'s and the nearest-neighbor (n.n) RKKY
interactions depend strongly on the local environment and are
overall {\it enhanced} by disorder. The disorder-induced
enhancement of $T_K$ in the clusters is in contrast to the
hypothesis of ``Kondo disorder'' models for extended systems.

The rest of the paper is organized as follows. In Sec. II, we
describe the model for both the periodic and disordered clusters.
In Secs. IIIA and IIIB we present results for the ground-state and
thermal properties of the ordered and disordered nanoclusters,
respectively. Section IV contains concluding remarks.

\section{METHODOLOGY}
The one dimensional Anderson lattice model is

\begin{eqnarray}
H =-t\sum_{i\sigma }(c_{i\sigma }^{\dagger }c_{i+1\sigma }+H.c)+
E_{f}\sum_{i\sigma }n_{i\sigma }^{\mathit{f}} \nonumber \\
+U\sum_{i}n_{i\uparrow }^{\mathit{f}}n_{i\downarrow }^{\mathit{f}%
} +V\sum_{i\sigma }(\mathit{f}_{i\sigma }^{\dagger }c_{i\sigma
}+H.c.).
\end{eqnarray}

\noindent Here, t is the nearest-neighbor hopping matrix element
for the conduction electrons, $c_{i,\sigma }^{+}(c_{i,\sigma })$
and $f_{i,\sigma }^{+}(f_{i,\sigma })$ create (annihilate) Wannier
electrons in $c$- and $f$- like orbital on site i with spin
$\sigma $, respectively; $E_{f}$ is the energy level of the bare
localized orbital, $V$ is the on-site hybridization matrix element
between the local f orbital and the conduction band and $U$ is the
on-site Coulomb repulsion of the f electrons. We use a simple
nearest-neighbor tight-binding model for the conduction band
dispersion, $ \epsilon _{k}=-2tcosk$. We consider the half-filled
($N_{el}=2N$) symmetric ($E_{f}=-\frac{U}{2}$) case, with $U = 5
(6)$ for the periodic (disorder) case. We investigate
one-dimensional rings of $N = 4$ and 6. Most of the results
presented are for the $N = 6$ case, except for the results for
$T>0$ where we have used $N=4$ sites. The exact diagonalization
calculations employ periodic boundary conditions.

\subsection{Ordered Clusters}

We have investigated the ground-state properties as a function of
the hybridization and the energy spacing in the conduction band,
$\Delta = 4t/(N-1) = \frac{4t}{5}$. We have calculated the
 average $f-$ and $c-$local moments, $<(\mu_i^f)^2>\equiv
 <S_i^{f,z}S_i^{f,z}>$
and $<(\mu_i^c)^2>\equiv<S_i^{c,z}S_i^{c,z}>$, respectively. Here,
$S_{i}^{f}$ and $ S_{i}^{c}$ are given by
\begin{equation}
S_{i}^{f}=\frac{1}{2}\sum_{\sigma ,\sigma ^{^{\prime }}}\tau
_{\sigma \sigma ^{^{\prime }}}f_{i\sigma }^{+}f_{i\sigma
^{^{\prime }}}
\end{equation}
and
\begin{equation}
S_{i}^{c}=\frac{1}{2}\sum_{\sigma ,\sigma ^{^{\prime }}}\tau
_{\sigma \sigma ^{^{\prime }}}c_{i\sigma }^{+}c_{i\sigma
^{^{\prime }}},
\end{equation}
where $\tau $ are the Pauli matrices.

We have also calculated the zero-temperature {\it f-f} and {\it
f-c} spin correlation functions (SCF) $<S_i^fS_{i+1}^f> \equiv
<g|S^{f,z}_iS^{f,z}_{i+1}|g>$ and $<S_i^fS_i^c> \equiv
<g|S^{f,z}_iS^{c,z}_i|g>$, respectively. Here, $|g>$ is the
many-body ground state and $S^{f,z}_i$ is the z-component of the
f-spin at site i. As expected, the cluster has a singlet ground
state ($S_g=0$ where $S_g$ is the ground-state spin) at half
filling. We compare the  onsite Kondo correlation function
$<S_i^fS_i^c>$ and the nearest-neighbor RKKY correlation function
$<S_i^fS_{i+1}^f>$ to assign a state to the Kondo or RKKY regimes,
in analogy with mean field treatments\cite{lacroix}. The spin
structure factor related to the equal-time $f-f$ spin correlation
functions in the ground state is
\begin{equation}
S^{ff}(q)=\frac{1}{N}\sum_{i,j}<g|{\bf S}_{i}^{f}\cdot {\bf
S}_{j}^{f}|g>e^{iq(x_i-x_j)}.
\end{equation}

The temperature-dependent local f-spin susceptibility,
$\chi^f(T)$,is
\begin{eqnarray}
\frac{ k_BT\chi^f(T )}{(g\mu_B)^2}= \frac{1}{Q} \sum_{\alpha}
e^{-\frac{E_{\alpha}}{k_BT}}<\alpha|S^f(i)S^{Tot}|\alpha>,
\end{eqnarray}
where
\begin{eqnarray}
Q = \sum_{\alpha} e^{-\frac{E_{\alpha}}{k_BT}}
\end{eqnarray}
\noindent is the partition function. Here, $S^{Tot}$ is the
z-projection of the total spin (both the $f$- and
$c$-contributions), and $|\alpha>$ and $E_{\alpha}$ are the exact
many-body eigenstates and eigenvalues, respectively. When $V=0$,
the localized spins and conduction electrons are decoupled and
$\chi^f(T)$ is simply the sum of the Curie term due to the free f
spins and the Pauli term of the free conduction electrons. For
finite $V$, $\chi^f(T)$ decreases with temperature at
low-temperatures. The specific heat is calculated from the second
derivative of the free energy $F$,
$C_v=-T\frac{\partial^{2}F}{\partial T^2}$. At $V=0$, the specific
heat of the system is given by the sum of the delta function at
$T=0$ that originates from the free localized spins and the
specific heat of free conduction electrons. For finite $V$ the
specific heat exhibits a double-peak structure: the
high-temperature peak is almost independent of the hybridization
and arises from the free conduction electron contribution,
whereas the low-temperature peak varies strongly with
hybridization.

\subsection{Disordered Clusters}
We consider a random binary alloy cluster, $A_{N-x}B_x$, of N=6
sites and different number of B atoms, $x = $ 0-N, arranged in a
ring described by the half-filled ($N_{el}=2N=12$) two-band
lattice Anderson Hamiltonian in Eq.(1). We introduce binary
disorder in the $f$-orbital energy $\epsilon^i_f$ ($=\epsilon_f^A$
or $\epsilon_f^B$) and in the intra-atomic Coulomb energy $U_i$
($=U^A$ or $U^B$), to model a Kondo-type A atom with $\epsilon_f^A
= - U^A/2$= -3 (symmetric case) and a mixed-valent (MV) type B
atom with $\epsilon_f^B$ = -2 and $U^B$ = 1. For both types of
atoms
 $V^A= V^B =V= 0.25$. For
$t=1$, this choice of parameters leads to a degeneracy between the
doubly-degenerate $c$-energy levels, $\epsilon_k = -t$, and the
energy level $\epsilon_f^B +U^B$. Upon filling the single particle
energy levels for any $x$, $N-x$ ($x$) electrons fill the
$\epsilon_f^A$ ($\epsilon_f^B$) levels, and two electrons the
-2$t$ conduction energy level, with the remaining $N-2$ electrons
accommodated in the $x$+4 degenerate states at $-t$. This in turn
results in strong charge fluctuations.

The temperature-dependent $f$ susceptibility, $\chi^f_x(T)$, at
concentration $x$, is given by
\begin{eqnarray}
\frac{ k_BT\chi^f_x(T )}{(g\mu_B)^2}= \frac{1}{Q}
\sum_{C_x,\alpha_{C_x}}
e^{-\frac{E_{\alpha_{C_x}}}{k_BT}}<\alpha_{C_x}|S^f(i)S^{Tot}|\alpha_{C_x}>,
\end{eqnarray}
\noindent where $C_x$ denote the configurations at concentration
$x$, $|\alpha_{C_x}>$ and $E_{C_x}$ are the
configuration-dependent exact many-body eigenstates and
eigenvalues, respectively, and $Q$ denotes the partition function.

\section{RESULTS AND DISCUSSION}
\subsection{Ordered Clusters}
\noindent {\it \underline{1. Ground State Properties}}\\

\indent In Fig. 1 we present the variation of the local Kondo SCF
$<S_i^fS_i^c>$ (squares) and the nearest-neighbor RKKY SCF
$<S_i^fS_{i+1}^f>$ (circles) as a function of hybridization for
two values of the hopping matrix element $t=0.2$ (closed symbols)
and $t=1.2$ (open symbols), respectively. As expected, for weak
hybridization V the local nearest-neighbor RKKY (Kondo) SCF is
large (small), indicating strong short-range antiferromagnetic
coupling between the
 $f-f$ local moments, which leads to long range magnetic ordering for
 extended systems. As V increases,
$<S_i^fS_{i+1}^f>$ decreases whereas the $<S_i^fS_i^c>$ increases
(in absolute value) saturating at large values of V. This gives
rise to the condensation of independent local Kondo singlets at
low temperatures, i.e., a disordered spin liquid phase. For large
$V$ the physics are {\it local}. Interestingly, as $t$ or $\Delta$
decreases the {\it f-c} spin correlation function is dramatically
enhanced while the {\it f-f} correlation function becomes weaker,
indicating a transition from the RKKY to the Kondo regime.
\begin{figure}[ht]
\hspace{2.in}{\includegraphics[width=.55\textwidth]{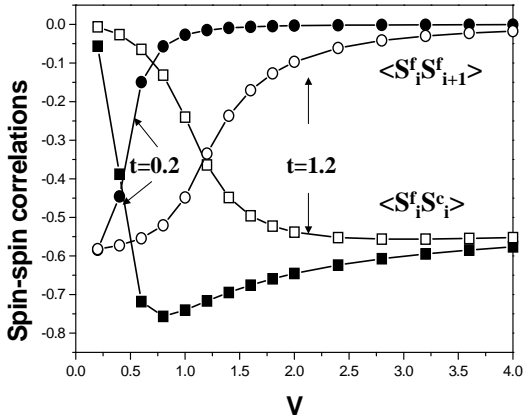}}
\vspace{-0.4 in}
 \caption{ Nearest neighbor f-f spin-spin
correlations (circles) and on-site f-c spin-spin correlations
(squares) as a function of V for two values of the hopping
parameter of $t=0.2$ (closed symbols) and $t=1.2$ (open symbols),
respectively.} \label{fig1}
\end{figure}
In Fig. 2 we present the average local $f$- (circles) and $c$-
(squares)  moments
 as a function of hybridization for
two values of the hopping matrix element $t=0.2$ (closed symbols)
and $t=1.2$ (open symbols), respectively.
 In the
weak hybridization limit, the large on-site Coulomb repulsion
reduces the double occupancy of the f level and a well-defined
local f moment is formed $\langle \mu_{f}^{2} \rangle = 1.0$ while
$\langle \mu_{c}^{2} \rangle = 0.5$. With increasing V both
charge- and spin- f1uctuations become enhanced and the local $f-$
moment decreases monotonically whereas the $c-$ local moment
exhibits a maximum. In the large V limit both the $f-$ and $c-$
local moments show similar dependence on V, with $<\mu_{c}^{2}>
\approx <\mu_{f}^{2}>$, indicating that the {\it total} local
moment $\mu$ vanishes. The effect of lowering the energy spacing
$\Delta$ is to decrease (increase) the $f-$ ($c-$) local moment,
thus tuning the magnetic behavior of the system. Note that the
 maximum value of the $c-$ local moment increases as $\Delta$ decreases.
This is due to the fact that for smaller $t$ values the kinetic
energy of conduction electrons is lowered, allowing conduction
electrons to be captured by f electrons to screen the local f
moment, thus leading to an enhancement of the local $c-$ moment.
\begin{figure}[ht]
\hspace{2.in}{\includegraphics[width=.55 \textwidth]{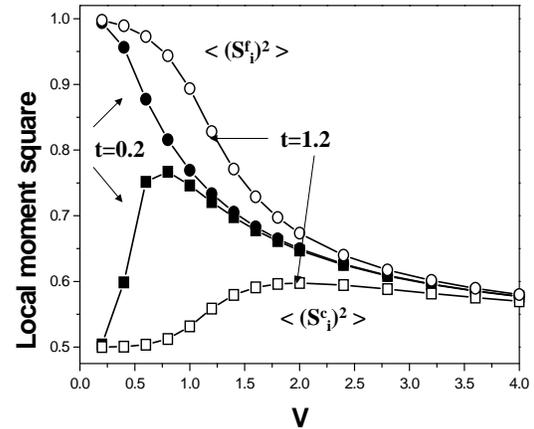}}
\vspace {-0.4 in}
 \caption{$f$- (circles) and $c-$ (squares) local
moment versus hybridization for two  values of the hopping
parameter of t=0.2 (closed symbols) and t=1.2 (open symbols),
respectively.} \label{fig2}
\end{figure}
In Fig. 3 we present the energy spacing versus V zero-temperature
phase diagram of the nanocluster, which illustrates the interplay
between Kondo and RKKY interactions. In the RKKY region
$<S_i^fS_{i+1}^f>$ is larger than the $<S_i^fS_i^c>$ and the {\it
total} local moment is non zero; in the Kondo regime
$<S_i^fS_{i+1}^f>$ is smaller than the $<S_i^fS_i^c>$, the {\it
total} local moment vanishes, and the ground state of the system
is composed of independent local singlets. The solid crossover
curve indicates the $V=V_c$ or $\Delta=\Delta_c$ values, where the
{\it local} and {\it non local} spin correlation functions are
equal, i.e., $<S_i^fS_{i+1}^f> = <S_i^fS_i^c>$. The dashed curve
denotes the set of points where the on-site {\it total} local
moment $\mu = 0$. Thus, in the intermediate regime, which will be
referred to as the {\it free spins} regime \cite{schroder},
$<S_i^fS_{i+1}^f>$ is smaller than the $ <S_i^fS_i^c>$, the $f$
moment is {\it partially} quenched and $\mu \not= 0$.
Interestingly, we find that
 the {\it free spins} regime becomes narrower as
the average level spacing $\Delta$ is reduced. This result may be
interpreted as a quantum critical regime (QCP) for the nanoring
due to the finite energy spacing, which eventually reduces to a
quantum critical point when $\Delta \rightarrow 0$.
\begin{figure}[ht]
\hspace{2.in}{\includegraphics[width=.55\textwidth]{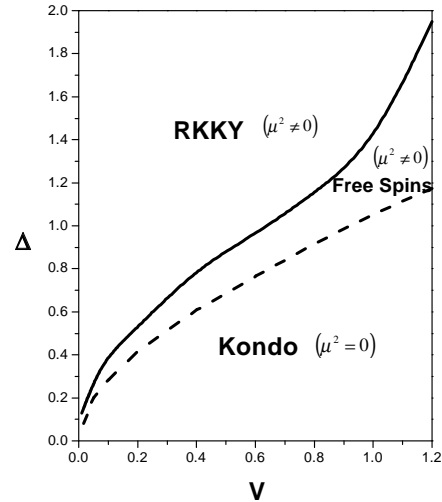}}
\vspace {-0.25in}
 \caption{Energy spacing $\Delta$ versus
hybridization zero-temperature phase diagram. The solid curve
denotes the crossover point of the spin-spin correlation function
in Fig.1; the dashed curve denotes the set of points where the
on-site total moment square $ \langle (\mu_{f}+\mu_{c})^{2}
\rangle = 0.0 \pm 0.05$.} \label{fig3}
\end{figure}
Fig. 4 shows the spin structure factor of the local f electrons
$S^{ff}(q)$ for various values of $V$ and for $t=0.2$. As
discussed earlier, the ground state of the half-filled symmetric
periodic Anderson model is a singlet. For small V, the spin
structure factor exhibits a maximum at $q=\pi$, indicating the
presence of strong antiferromagnetic correlations between the
local f moments, consistent with the large values of
$<S_i^fS_{i+1}^f>$ in Fig. 1. With increasing hybridization, the
maximum of $S^{ff}(q=\pi)$ decreases and vanishes at very large
hybridization, indicating that the ground state undergoes a
transition from the antiferromagnetic to the nonmagnetic Fermi
liquid phase. This is consistent with the zero-temperature phase
diagram in Fig. 3.
\begin{figure}[ht]
\hspace{2.in}{\includegraphics[width=.55\textwidth]{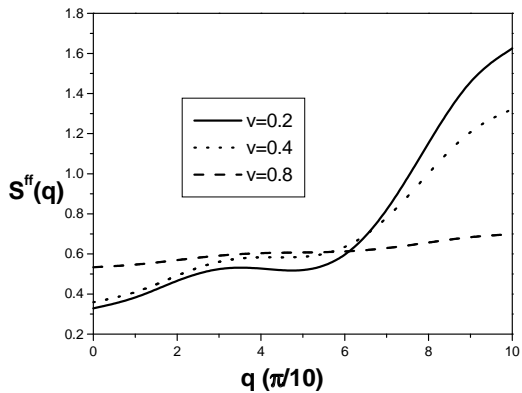}}
\vspace{-0.7 in}
 \caption{Spin structure factor as a function of
wave-vector for different values of $V$ and for t=0.2. }
\label{fig4}
\end{figure}
The spin gap as a function of hybridization $V$ for two values of
energy spacing is shown in Fig. 5.  The spin gap is defined as the
energy difference between the singlet ground state and the
lowest-lying excited triplet $(S=1)$ state. As expected, there is
a nonzero spin gap for the half-filled Anderson lattice model,
which increases with hybridization. Interestingly, the spin gap
dramatically increases as the average energy level spacing
$\Delta$ is reduced. Thus, the energy spacing or equivalently the
size of the cluster tunes the low-energy excitation energy  which
controls the low-temperature specific heat and susceptibility.
\begin{figure}[ht]
\hspace{2.in}{\includegraphics[width=.55\textwidth]{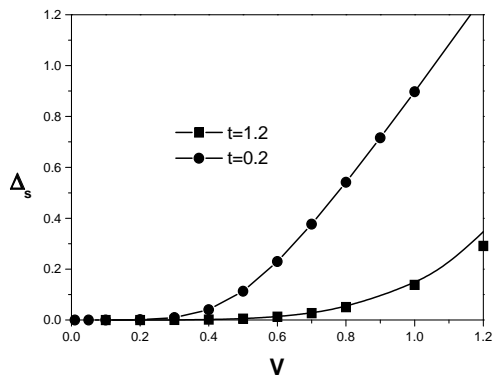}}
\vspace {-0.8 in} \caption{Spin gap as a function of $V$ for
$t=0.2$ and 1.2. The spin gap increases exponentially (linearly)
for small (large) V.} \label{fig5}
\end{figure}
\vspace{ 0.1 in}

\noindent {\it \underline{2. Thermal
Properties}}\\
\\
The T=0 exact diagonalization results on small clusters are
generally plagued by strong finite size
effects\cite{Pastor,haule}. Performing calculations at $T>0$ gives
not only the thermodynamic properties of the system, but most
importantly diminishes finite-size effects for $(k_{B}T \gg
\Delta)$.

In Fig. 6, we show the nearest-neighbor f-f spin-spin correlations
and on-site f-c spin-spin correlation as a function of temperature
for for $t=0.2$ and for $V= 0.2 < V_c$ and $V= 0.4
> V_c$, where $V_c=0.25$. At high temperatures, the free
moments of the f and conduction electrons are essentially
decoupled. The nearest-neighbor {\it non local} spin correlation
function falls more rapidly with $T$ than the on-site {\it local}
$f-c$ spin-spin correlations, indicating that the {\it non local}
spin correlations can  be destroyed easier by thermal
fluctuations. For $V<V_c$, the nanocluster is dominated by RKKY
(Kondo) interactions at temperatures lower (higher) than the
crossover temperature, $T^{cl}_{RKKY}$, which denotes the
temperature where the {\it non local} and {\it local} interaction
become equal in the nanocluster. In the infinite system this
temperature would denote the ordering N$\acute{e}$el temperature.
On the other hand, for $V>V_c$ the RKKY and Kondo spin correlation
functions do not intersect at any $T$, and the physics become
dominated by the local interactions.
\begin{figure}[ht]
\hspace{2.in}{\includegraphics[width=.55\textwidth]{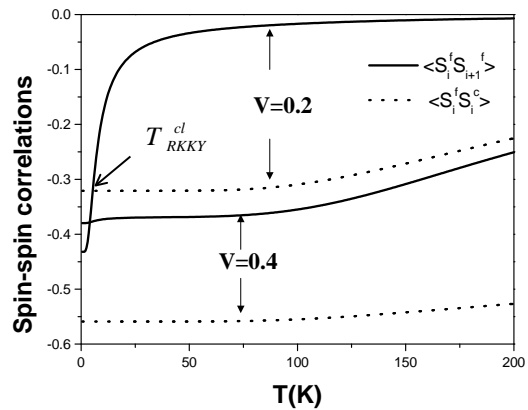}}
\vspace {-0.5 in} \caption{Nearest-neighbor f-f and on-site f-c
spin-spin correlation functions versus temperature for $t=0.2$ and
for $V=0.2<V_c$ and $V=0.4>V_c$, $V_c=0.25$.} \label{fig6}
\end{figure}
In Fig. 7 we present the crossover temperature $T^{cl}_{RKKY}$ for
the cluster as a function of hybridization for different values of
$t$. This represents the phase diagram of the strongly correlated
nanocluster, which is similar to the ``Doniach phase diagram" for
the infinite Kondo necklace model. The phase within the crossover
curve denotes the regime where the {\it non local} short-range
magnetic correlations are dominant. For $V<V_c$ and
$T>>T^{cl}_{RKKY}$ one enters into the disordered ``free" local
moment regime. On the other hand, for $V>V_c$ and at low $T$, the
 nanocluster can be viewed as a condensate of singlets,
typical of the Kondo spin-liquid regime.  Interestingly, the
$T^{cl}_{RKKY}$ can be tuned by the energy spacing $\Delta$ or the
size of the cluster. Thus, increasing $\Delta$ or decreasing the
size of the nanocluster results to enhancement of the {\it non
local} nearest-neighbor magnetic correlations and hence
$T^{cl}_{RKKY}$. This result is the first exact ``Doniach phase
diagram" for a nanocluster.
\begin{figure}[ht]
\hspace{2.in}{\includegraphics[width=.55\textwidth]{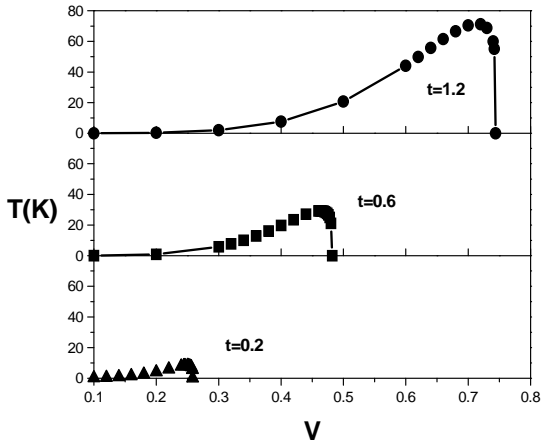}}
\vspace {-0.4 in} \caption{Effect of energy spacing,
$\Delta=\frac{4t}{N-1}$ on the exact ``Doniach phase diagram" for
a strongly correlated electron nanochain. The crossover curve
represents the  crossover temperature $T^{cl}_{RKKY}$, where the
non local short range AF spin correlations become equal to the
local on-site Kondo spin correlations.} \label{fig7}
\end{figure}

In bulk Kondo insulators and heavy-fermion systems, the low-$T$
 susceptibility  and specific heat behavior is determined by the spin
 gap, which for the half-filled Anderson lattice model,
 is determined by the ratio of $V$ to $U$.
 On the other hand, strongly correlated nanoclusters
 are inherently associated with a new low-energy cutoff, namely
the energy spacing $\Delta$ of the conduction electrons. Thus, a
key question is how can the low-temperature physics be tuned by
the interplay of the spin gap and the energy spacing. In Fig. 8 we
present the local f magnetic susceptibility as a function of
temperature for $t=0.2$ and for $V= 0.2 < V_c$, $V=V_c=0.25$, and
$V=0.4 > V_c$. For small $V$, the spin gap which is smaller than
$\Delta$ controls the exponential activation behavior of $\chi^f$
at low $T$. On the other hand, in the large $V$ limit, the spin
gap becomes larger than $\Delta$ (see Fig. 5) and the low-$T$
behavior of the susceptibility shows no exponential activation. At
high $T$ we can see an asymptotic Curie-Weiss regime, typical of
localized decoupled moments.
\begin{figure}[ht]
\hspace{2.in}{\includegraphics[width=.55\textwidth]{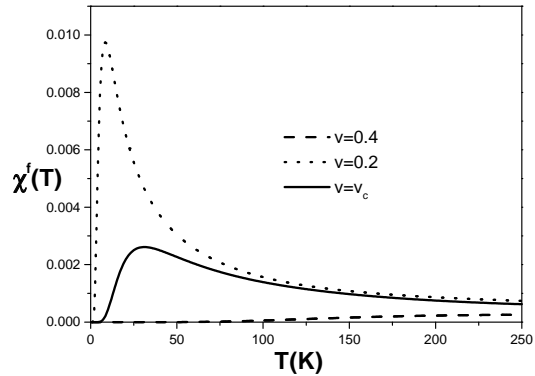}}
\vspace {-0.6 in} \caption{Local f magnetic susceptibility as a
function of temperature for $t=0.2$ and with $V=0.2 < V_c$, $V =
V_c = 0.25$ and $V=0.4>V_c$.} \label{fig8}
\end{figure}
In Fig. 9, we present the specific heat as a function of
temperature for $V=0.4$ and different $t$. At $V=0$, the specific
heat is given by the sum of a delta function at $T=0$ for the
localized spins and the specific heat of free fermions. As
expected, by switching on the coupling $V$, they are combined to
form a two-peak structure. The broad peak at high $T$ is rather
similar to the free-electron gas. The low-$T$ behavior is
associated with the lowest energy scale, which as in the case of
the susceptibility, is determined by the lowest value between the
spin gap and the energy spacing $\Delta$. For large values of $t$
(or $\Delta$) the spin gap is reduced (see Fig. 5) and the spin
gap is the lowest energy scale. Consequently, the low-$T$ behavior
exhibits exponential activation associated with the spin gap. On
the other hand, for small energy spacing the physics become local
(Kondo regime) and the low-$T$ sharp peak shifts towards higher
temperatures and becomes broader.
\begin{figure}[ht]
\hspace{2.in}{\includegraphics[width=.55\textwidth]{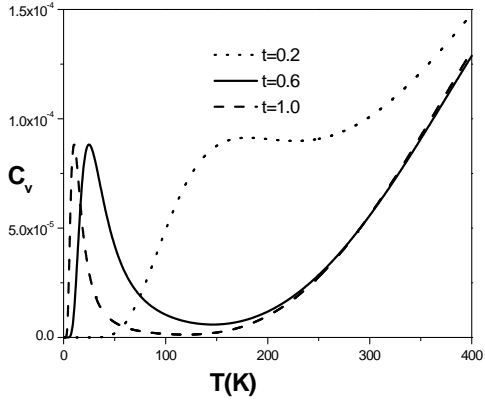}}
\vspace {-0.5 in} \caption{Specific heat as a function of
temperature with $V=0.4$ and various values of $t=$ 0.2, 0.6 and
1.0. The low-$T$ peak for larger energy spacing is due to the spin
gap.} \label{fig9}
\end{figure}
\subsection{Disordered Clusters}
\noindent {\it \underline{1. Effect of Disorder}}\\

The configurations for $x\leq3$ are shown in Fig. 10, left panel,
along with the value of the spin, S$_g$, of the ground-state. The
A (B) atoms are denoted by closed (open) circles, respectively.
Except for the homogenous cases ($x$=0 and $x=6$), with a S$_g$ =
0 ground state, for all $x$ there are configurations with S$_g
\neq 0$. The average occupation and average LM for the periodic
Kondo and MV lattices are $<n_f^A> = 1$, $<(\mu_f^A)^2>$ = 0.99,
and $<n_f^B>$ = 1.6, $<(\mu_f^B)^2>$ = 0.43, respectively. We
carry out a detailed analysis  for $x$=1 (S$_g$ =2) to demonstrate
the FM transition induced by a single MV atom in an otherwise
Kondo cluster. Studies of extended systems have reported similar
occurrence of ferromagnetism in the MV phase\cite{Nolting}. As
expected, the singlet ground state of the $x=0$ Kondo cluster is
characterized by n.n. anti-ferromagnetic (AF) {\it f-f} spin
correlations ($<S_f^A(i)S_f^A(i+1)>$ = - 0.58). The introduction
of a MV atom renders them ferromagnetic. Since $U_B$ is small, the
B impurity tends to remove charge from the the conduction band, in
particular from the $k$-state with $\epsilon_k = -t$, which has
large amplitude at the B site and at the opposite A site across
the ring.  Such a depletion is different for the two spin states,
thus yielding a maximum value for the f-moment of the MV atom. The
$f$-$f$ spin correlation function between the Kondo and MV atoms
are AF ($<S_f^A(i)S_f^B(i+1)>$ = - 0.23), while they are FM among
the Kondo atoms ($<S_f^A(i)S_f^A(i+1)>$ = +0.94). A similar result
was  recently found in {\it ab initio} calculations\cite {MnPRL},
where introducing a nitrogen impurity in small (1-5 atoms) Mn
clusters induces ferromagnetism via AF coupling between the N to
the Mn atoms, whilst Mn-Mn couple ferromagnetically. We find that
there is a crossover in $S_g$ from 0 $\rightarrow 1 \rightarrow  2
\rightarrow 0$ (Fig. 10, right panel) indicating a reentrant
nonmagnetic transition around $\epsilon_B=2$. This almost
saturated FM $S_g=2$ domain is robust against small changes in
$U_B$, $V$, $\epsilon_A$, $U_A$, cluster size ($N=7$), and band
filling ($N_{el}$ = 10) provided that the Kondo atom has a large
LM.
\begin{figure}[tb]
\hspace{2.in}{\includegraphics[width=.65\textwidth]{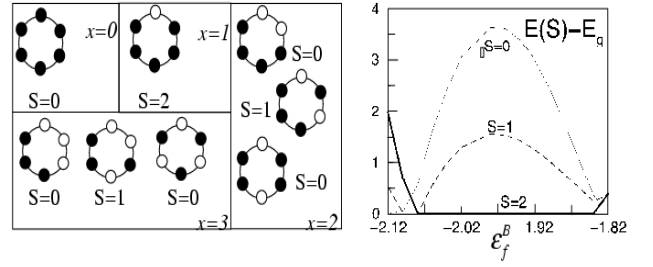}}
\vspace {-2.2in} \caption{\label{fig:epsart} Left panel: Alloy
configurations for various concentrations $x\leq 3$ (the $x>3$
cases are obtained by exchanging closed and open circles). For
each $x\leq3$ configuration, the value of the ground-state spin
$S_g$ is reported. Right panel: Energy difference (in units of
$10^{-4}t$ ) between the lowest $S\leq2$ eigenstates and the
ground state as function of $\epsilon_B$.}
\end{figure}

\begin{figure}[ht]
\centerline{\includegraphics[width=.5\textwidth]{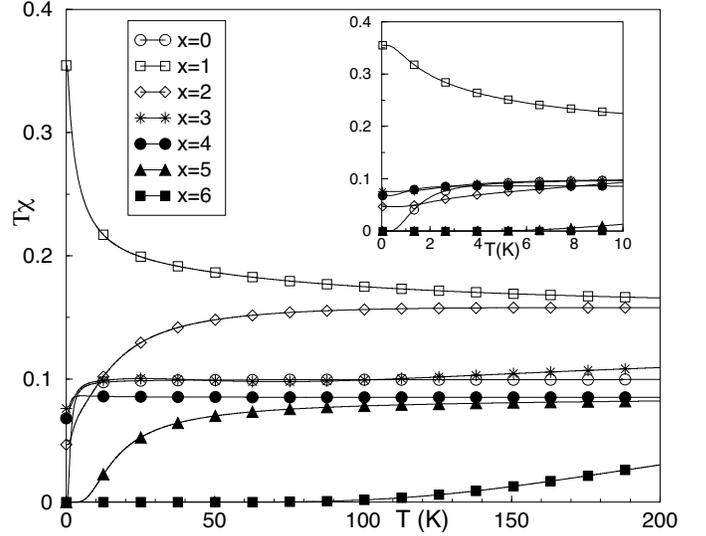}}
\caption{\label{fig:epsart2} Temperature dependence of the average
f-susceptibility for different alloy concentrations. The inset
shows the low-temperature behavior.} \protect\label{show.fig}
\end{figure}
In Fig. 11 we present $T\chi^f_x(T)$ as a function of temperature
for different $x$. As $T\rightarrow 0$ (inset Fig. 11)
$T\chi^f_x(T)$ approaches a finite value for $x=1-4$ while it
vanishes exponentially for $x$=0, 5 and 6. This is due to the fact
that the former concentrations involve some configurations which
are magnetic, while the latter have singlet ground states (Fig.
10). The stronger (weaker) low-temperature dependence for $x=1$
($x=2-4$) is due to the smaller (larger) spin gap between the
ground state and the lowest excited states. The magnetic
susceptibility displays also a magnetic crossover upon varying
$x$, and reveals a Curie-like divergence at low T for $x=1-4$. The
temperature-dependent results for the specific heat, not reported
here, show corroborative evidence of this disorder-induced
magnetic crossover.

\vspace{0.2 in}
 \noindent {\it \underline{2.
Effect of Energy Spacing}}\\

\noindent Next we address a number of important open issues,
namely
 (1) the effect of  $\Delta$ on the interplay between
 RKKY and Kondo interactions in disordered
 clusters, (2) the characterization of  the single-impurity "Kondo
correlation energy" $T_K$ in a {\it dense-impurity} cluster and
(3) the effect of disorder and $\Delta$ on the distribution of the
local $T_K$'s.  In the following, $\epsilon_B=-2$.

In contrast with previous studies, which introduced a
phenomenological distribution $P(T_K)$ of single-impurity Kondo
temperatures, the advantage of the present calculations is that
one calculates exactly the Kondo correlation energy: we employ the
so-called ``hybridization'' approach\cite{fulde}, with $T_K$
defined as
\begin{equation}
k_B T_K(i) = E_g(V_i=0)-E_g,
\end{equation}
where $E_g(V_i=0)$ is the ground-state energy of the
dense-impurity cluster when V is set to zero at the i{\it th}
site. Eq.(8) reduces to $ k_BT_K = E_{band}-E_F+\epsilon_f- E_g
$\cite{Hu,Yoshida} in the single impurity case. Here, $E_F$ is the
highest occupied energy level in the conduction band and
$E_{band}$ is the conduction band energy. This definition of the
local $T_K$ takes into account the interaction of the $f$-moment
at site $i$ with the other $f$-moments in the system
\cite{Wilkins}.
%


In Table I  we list for the periodic, $x$=0, case the local Kondo
{\it f-c} spin correlation function $<S_f^A(i)S_c^A(i)>$, the n.n.
{\it f-f} spin correlation function $<S_f^A(i)S_f^A(i+1)>$, and
the local Kondo temperature for two different values of $t$ (The
energy spacing is $\Delta=4t/(N-1)\equiv 4t/5$). As $t$ or
$\Delta$ decreases the {\it f-c} spin correlation function is
dramatically enhanced while the {\it f-f} correlation function
becomes weaker, indicating a transition from the RKKY to the Kondo
regime. This is also corroborated by the increase in the local
T$_K(i)$. The energy spacing affects not only the magnetic (A)
atoms but the MV atoms as well. Thus, increasing $t$  drives the B
atoms from the non-magnetic, NM ($n_f \approx 2$), to the MV  and
finally to the Kondo regime.

We next examine the role of even versus odd number of electrons on
the magnetic behavior of the uniform x=0 case. For $t=1$, changing
the number of electrons from N$_{el}$ = 12 to N$_{el}$ = 11
results in: (a) an enhancement of the local Kondo f-c spin
correlation function, $<S^A_f(i)S^A_c(i)>$ from  -0.01 to -0.12;
and (b) a suppression of the nearest-neighbor f-f spin correlation
function $<S^A_f(i)S^A_f(i+1)>$ from -0.58 to  -0.20 (Due to the
broken symmetry for N$_{el}$ = 11, the f-f spin correlation
functions range from -0.5 to +0.02). This interesting novel tuning
of the magnetic behavior can be understood as follows:  For an
odd-electron cluster, the topmost occupied single particle energy
level is singly occupied. On the other hand, for an even-electron
cluster, the topmost occupied single-particle energy level is
doubly occupied, thus blocking energy-lowering spin-flip
transitions.  This energy penalty intrinsically weakens the Kondo
correlations\cite{thimm}.  As expected, changing the number of
electrons from even to odd changes $S_g$ = 0 to $S_g=
\frac{1}{2}$. For t= 0.05 (Kondo regime), the on site f-c
correlation function does not depend as strongly on the parity in
the number of electrons because the sites are locked into
singlets. On the other hand, $<S^A_f(i)S^A_f(i+1)>$ becomes
suppressed as in the case of large energy spacing. Similar results
were found for the various disordered concentrations.
\begin{figure}[ht]
\centerline{\includegraphics[width=.5\textwidth]{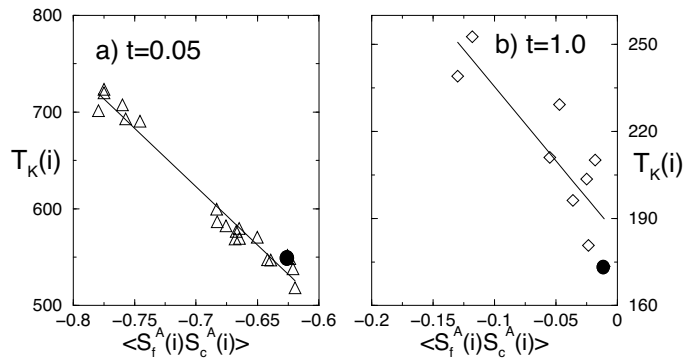}}
\caption{\label{fig:epsart} A-atoms: Local Kondo Temperatures (in
K) vs the local f-c spin correlation function, for different
configurations and two different values of $t$. The closed circles
refer to the $x=0$ case and the lines are a guide to the eye. }
\end{figure}

\begin{table}[ht]
\caption{\label{tab:table1}  Local Kondo  f-c and n.n. f-f spin
correlations functions and the local Kondo temperature (in K) for
two values of $t$ (in eV). The average energy spacing is
$\Delta=4t/(N-1)\equiv 4t/5$.}

\begin{tabular}{c|ccc}
 & $<S_f^A(i)S_c^A(i)>$ & $<S_f^A(i)S_f^A(i+1)>$ & $T_K(i) $ \\
\hline
 t=0.05  & -0.626  & -0.322  & 551.8  \\
\hline
 t=1.00  & -0.011  & -0.584  & 173.4  \\
\end{tabular}

\end{table}
In Fig. 12 we present the local $T_K(i)$ as a function of the
local {\it f-c} spin correlation function  $<S_f^A(i)S_c^A(i)>$
for all Kondo (A) atoms in the singlet ground state at any
concentration $x$ for $t$= 0.05 and 1.0. Note the different scales
both on the horizontal and vertical axis in the panels. In both
panels, the closed circles correspond to the $x$=0 lattice case
and the line is a guide to the eye. The results indicate a
correlation between T$_K$ and the {\it f-c} spin correlation
function (the larger $T_K$'s correspond to the more negative {\it
f-c} values) as one would expect, since both provide a measure of
the Kondo effect. For $t$=0.05, most of the disordered cluster
configurations are in the Kondo regime ($S_g$ = 0), with larger
$T_K$ values; consequently, panel (a) has a larger number of
singlet configurations. The introduction of MV impurities induces
a distribution of $T_K(i)$'s, whose values are overall {\it
enhanced} compared to those for the $x$=0 case, except for several
configurations for $t$=0.05, in contrast with single-site theories
for extended systems\cite{Miranda}. It is interesting that
$P(T_K)$ for $t$=0.05 exhibits a bimodal behavior centered about
710 and 570K, respectively: The higher $T_K$'s originate from
isolated Kondo atoms which have MV atoms as n.n. so that the local
screening of the magnetic moment of the A atom is enhanced.
%

The effect of alloying  and $\Delta$ on the RKKY versus Kondo
competition for a given $x$ is seen in Fig. 13 (left panel), where
the configuration averaged local $<S_f^A(i)S_c^A(i)>_x$ and
$<S_f^A(i)S_f^A(i+1)>_x$ correlation functions are plotted as a
function of $t$. The solid curves denote the uniform $x$=0 case,
where we drive the cluster from the RKKY to the Kondo regime as we
decrease $t$. We find that the stronger the average Kondo
correlations are the weaker the average RKKY interactions and vice
versa. In the weak Kondo regime the configurations exhibit a wider
distribution of RKKY interactions indicating that they are
sensitive to the local environment. In contrast, in the strong
Kondo regime, the Kondo (A) atoms become locked into local Kondo
singlets and the n.n. RKKY interactions are insensitive to the
local environment. Interestingly, both energy spacing and disorder
lead to an overall enhancement of the RKKY interactions compared
to the homogenous state.
\begin{figure}[ht]
\hspace{2.in}{\includegraphics[width=.6 \textwidth]{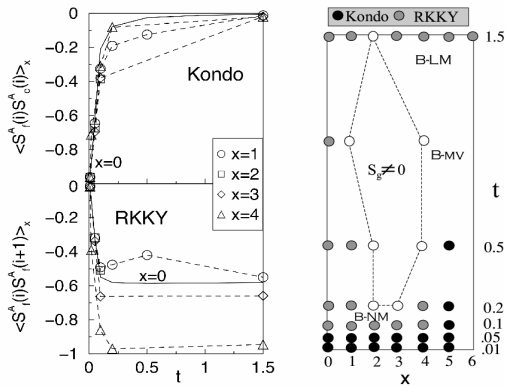}}
\vspace{-3.2cm}
 \caption{\label{fig:epsart} Left panel:
Configuration-averaged local f-c (top) and n.n. f-f spin
correlations (bottom) for the A atoms as function of $t$. The
solid line refers to the homogenous $x=0$ case. Right panel:
Zero-temperature $t$ vs $x$ phase diagram for the nanocluster.
Black (gray) circles denote the Kondo (RKKY) regime. The white
circles and the dashed contour delimit the FM region. The
horizontal stripes denote the non-magnetic (NM), mixed valence
(MV)  and local moment (LM) behavior of the B-atoms. }
\end{figure}
In the right panel of Fig. 13 we present the  $t$ versus $x$ phase
diagram for the nanocluster at $T=0$ . We compare the
$<S_f^A(i)S_c^A(i)>_x$  and  $<S_f^A(i)S_f^A(i+1)>_x$ to assign a
state of specific concentration to the Kondo or RKKY regimes
(black and gray circles, respectively), in analogy with the $x=0$
case (Table I) and with mean field treatments\cite{lacroix}. The
horizontal gray stripes denote qualitatively ranges of $t$ where
the B atoms exhibit NM, MV and LM behavior. An interesting feature
of the phase diagram is the appearance of a large FM region ($S_g
\neq 0 $) enclosed by the dashed line. The RKKY region at large
$t$ and large $x$ originates from the B atoms which become
magnetic. For the non FM configurations and for $x<5$ the  Kondo
(RKKY) correlations of the A atoms dominate at small (large) $t$,
in analogy with the $x=0$ case. On the other hand, for $x=5$ the
local Kondo correlations of the single A atom at low $t$ dominate
over the {\it f-f} correlations between the A-B and B-B pairs. For
the uniform ($x$=6) MV case we include only results in the large
$t$ regime, where the MV atoms acquire LM's which couple
antiferromagnetically. Overall, the RKKY interactions prevail for
any concentration when $t$ is comparable or larger than the
hybridization $V$.

\section{CONCLUSIONS}

Recent advances in STM experiments have made it possible to study
the electronic and magnetic properties of strongly correlated
electrons in nanoscopic and mesoscopic systems. There are two main
differences between nanosized clusters and the infinite lattice.
First, the discrete energy levels of the conduction band states
introduce a new low-energy scale, i.e., the average energy level
spacing $\Delta$. This new energy scale that competes with the
spin gap can effect the low-temperature behavior of the system.
Second, the results depend on the parity of the total number of
electrons. If $N_{tot}$ is odd, the ground state is doubly
degenerate.

We have carried out exact diagonalization calculations which
reveal that the: (1) energy spacing;  (2) parity of the number of
electrons; and (3) disorder, give rise to a novel tuning of the
magnetic behavior of a {\it dense} Kondo nanocluster.  This
interesting and important tuning can drive the nanocluster from
the Kondo to the RKKY regime, i.e. a tunable ``Doniach" phase
diagram in nanoclusters. We have employed the criterion of
comparing the exact {\it non local} versus {\it local} spin
correlation functions to determine if the nanocluster lies in the
RKKY versus Kondo regime. For weak hybridization, where the spin
gap is smaller than $\Delta$, both the low-temperature local $f$
susceptibility and specific heat exhibit an exponential activation
behavior associated with the spin gap. In contrast in the large
hybridization limit, $\Delta$ is smaller than the spin gap, the
physics become local and the exponential activation behavior
disappears. The interplay of $\Delta$ and disorder produces a rich
structure zero-temperature alloy phase diagram, where regions with
prevailing Kondo or RKKY correlations alternate with domains of FM
order. The distribution of local $T_K$ and RKKY interactions
depends strongly on the local environment and are overall {\it
enhanced} by disorder, in contrast to the hypothesis of
single-impurity based ``Kondo disorder'' models for extended
systems. We believe that the conclusions of our calculations
should be relevant to experimental realizations of small clusters
and quantum dots. For example, the recent experiments\cite{odom}
of magnetic clusters on single-walled carbon nanotubes of varying
size provide much flexibility for investigating the interplay of
Kondo and RKKY effects at different energy scales.

\begin{acknowledgments}
The research at California State University Northridge was
supported through NSF under Grant Nos. DMR-0097187, NASA under
grant No. NCC5-513, and the Keck and Parsons Foundations grants.
The calculations were performed on the the CSUN Massively Parallel
Computer Platform supported through NSF under Grand No.
DMR-0011656. We acknowledge useful discussions with P. Fulde, P.
Schlottmann, P. Riseborough, A.H. Castro Neto, P.Cornaglia and C.
Balseiro.
\end{acknowledgments}


\begin{thebibliography}{99}
\bibitem{Hewson} A.C.Hewson,{\it The Kondo Problem to Heavy Fermions}, Cambridge Press, New York, 1993.
\bibitem{1960} P.W.Anderson, Phys.Rev. {\bf124} ,41 (1961); J.Kondo, Progr. Theor. Phys.  {\bf32} , 37 (1964).
\bibitem{Steglich} C.D.Bredl, S.Horn, F.Steglich, B.Luthi and R.M.Martin, Phys. Rev. Lett.  {\bf52}, 1982 (1984).
\bibitem{UedaRMP} H.Tsunetsugu, M. Sigrist and K.Ueda, Rev.Mod.Phys.  {\bf69} , 809 (1997).
\bibitem{Miranda} E.Miranda, V.Dobrosavljevic and G.Kotliar, Phys. Rev. Lett.  {\bf78}, 290 (1997).
\bibitem{CastroNeto} A.H.Castro Neto and B.A. Jones, Phys.Rev. B  {\bf62}, 14975 (2000).
\bibitem{Stewart} G.R.Stewart, Rev. Mod. Phys.  {\bf 73}, 797 (2001).
\bibitem{Schlottman} P.Schlottmann, Phys. Rev. B  {\bf65}, 174407 (2002).
\bibitem{Si} J. Lleweilun Smith and and Q. Si, Phys. Rev. B {\bf 61},
5184 (2003).
\bibitem{doniach} S.Doniach, Physica B {\bf 91}, 231 (1977).
\bibitem{schroder}A. Schr$\ddot{o}$der, G. Aeppli, R. Coldea, M. Adams,
O. Stockert, H.V. L$\ddot{o}$hneysen, E. Bucher, R. Ramazashvili
and P. Coleman, Nature {\bf 407}, 351 (2000).
\bibitem{Riseborough} P.S.Riseborough, Phys.Rev.B  {\bf 45}, 13984 (1992).
\bibitem{Bernal} O.O.Bernal, D.E. Maclaughlin, H.G.Lukefahr and B.Andraka, Phys. Rev. Lett. {\bf 75}, 2023 (1995).
\bibitem{rice} T.M.Rice and K.Ueda, Phys. Rev. Lett. {\bf 55}, 995 (1985).
\bibitem{read} N.Read, D.M.Newns, and S.Doniach, Phys. Rev. B{\bf 30}, 3841 (1984); S.Burdin, A.Georges, and D.R.Grempel, Phys. Rev. Lett. {\bf 85}, 1048 (2000).
\bibitem{assaad}S.Capponi and F.F.Assaad, Phys. Rev. B {\bf 63},155114 (2001).
\bibitem{Miranda2001}  E.Miranda and V. Dobrosavljevic, Phys. Rev. Lett. {\bf 86} 264 (2001).
\bibitem{gordon} D.G.Gordon, H.Shtrikman, D. Mahalu, D.A. Magder, U.Meirav, and M.A.Kaster, Nature (London) {\bf 391}, 156 (1998).
\bibitem{thimm} W.B. Thimm, J. Kroha, and J.V. Delft, Phys. Rev. Lett. {\bf 82}, 2143 (1999).
\bibitem{manoharan} H.C. Manoharan, C.P. Lutz, and D.M. Eigler, Nature (London) {\bf 403}, 512 (2000).
\bibitem{odom} T. Odom, J.L. Huang, C. Li Cheung, and C. M. Lieber, Science {\bf 290}, 1549 (2000) and references therein.
\bibitem{Hu} H.Hu, G.M. Zhang and L. Yu, Phys. Rev. Lett.  {\bf 86}, 5558 (2001).
\bibitem{Balseiro} P.S. Cornaglia and C.A. Balseiro, Phys.Rev. B  {\bf 66},115303 (2002).
\bibitem{Affleck} P. Simon and I. Affleck, Phys.Rev.Lett  {\bf 89},206602,(2002).
\bibitem{schlottmann2001} P. Schlottmann, Phys.Rev.B {\bf 65}, 024431 (2001).
\bibitem{Pastor} G.M.Pastor, R.Hirsch and B.M\"{u}hlschlegel, Phys.Rev.Lett.
{\bf 72}, 3879 (1994).
\bibitem{lacroix}B.Coqblin, C. Lacroix, M.S. Gusmao and J.R. Iglesias,
Phys.Rev. B  {\bf 67},064417(2003).
\bibitem{haule} K.Haule, J.Bonca and P. Prelovsek, Phys.Rev.B {\bf 61}, 2482 (2000).
\bibitem{Nolting} D.Meyer and W.Nolting, Phys.Rev.B  {\bf 62}, 5657 (2000); D.Meyer, Solid State Comm.  {\bf 121}, 565 (2002).
\bibitem{MnPRL} B.K.Rao and P.Jena , Phys.Rev.Lett.  {\bf 89}, 185504 (2002).
\bibitem{fulde} P.Fulde, private communication, and P.Fulde, "Electron Correlations in Molecules and Solids", 3rd edition, Springer-Berlin-(1995).
\bibitem{Yoshida} K.Yosida, Phys.Rev.  {\bf 147}, 223 (1966).
\bibitem{Wilkins} We also employed a second approach,
$k_BT'_K = \mu_iB_c$, where $B_c$ is the critical local external
magnetic field  necessary to break up the singlet bound state
\cite{Balseiro}. Comparative results of the methods will be
presented elsewhere.
\end{thebibliography}
\end{document}